\documentclass[preprint]{vgtc}               

\ifpdf
  \pdfoutput=1\relax                   
  \usepackage{graphicx}                
  \DeclareGraphicsExtensions{.pdf,.png,.jpg,.jpeg} 
\else
  \usepackage{graphicx}                
  \DeclareGraphicsExtensions{.eps}     
\fi%

\graphicspath{{figures/}{pictures/}{images/}{./}} 

\usepackage{microtype}                 
\PassOptionsToPackage{warn}{textcomp}  
\usepackage{textcomp}                  
\usepackage{mathptmx}                  
\usepackage{times}                     
\usepackage{cite}                      
\usepackage{tabu}                      
\usepackage{booktabs}  
\usepackage{tabularx}
\usepackage{gensymb}
\makeatletter
\renewcommand\paragraph{\@startsection{paragraph}{4}{\z@}%
          {-2.5ex \@plus -1ex \@minus -.2ex}%
          {1.5ex \@plus .2ex}%
          {\normalfont\normalsize}}
\makeatother

\onlineid{0}

\vgtccategory{Research}

\vgtcinsertpkg

\preprinttext{This is the author's version, to appear in the IEEE VR 2025 conference.}

\title{Exploring the Effects of Level of Control in the Initialization of Shared Whiteboarding Sessions in Collaborative Augmented Reality}

\author{Logan Lane $^1$ \thanks{e-mail: logantl@vt.edu} %
\and Jerald Thomas $^2$
\and Alexander Giovannelli $^1$
\and Ibrahim Tahmid $^1$
\and Doug A. Bowman $^1$}
\affiliation{
    \scriptsize $^1$ Center for Human-Computer Interaction, Virginia Tech, USA \\ \scriptsize $^2$ University of Wisconsin Milwaukee, USA %
}

\teaser{
  \centering
  \includegraphics[width=.9\linewidth]{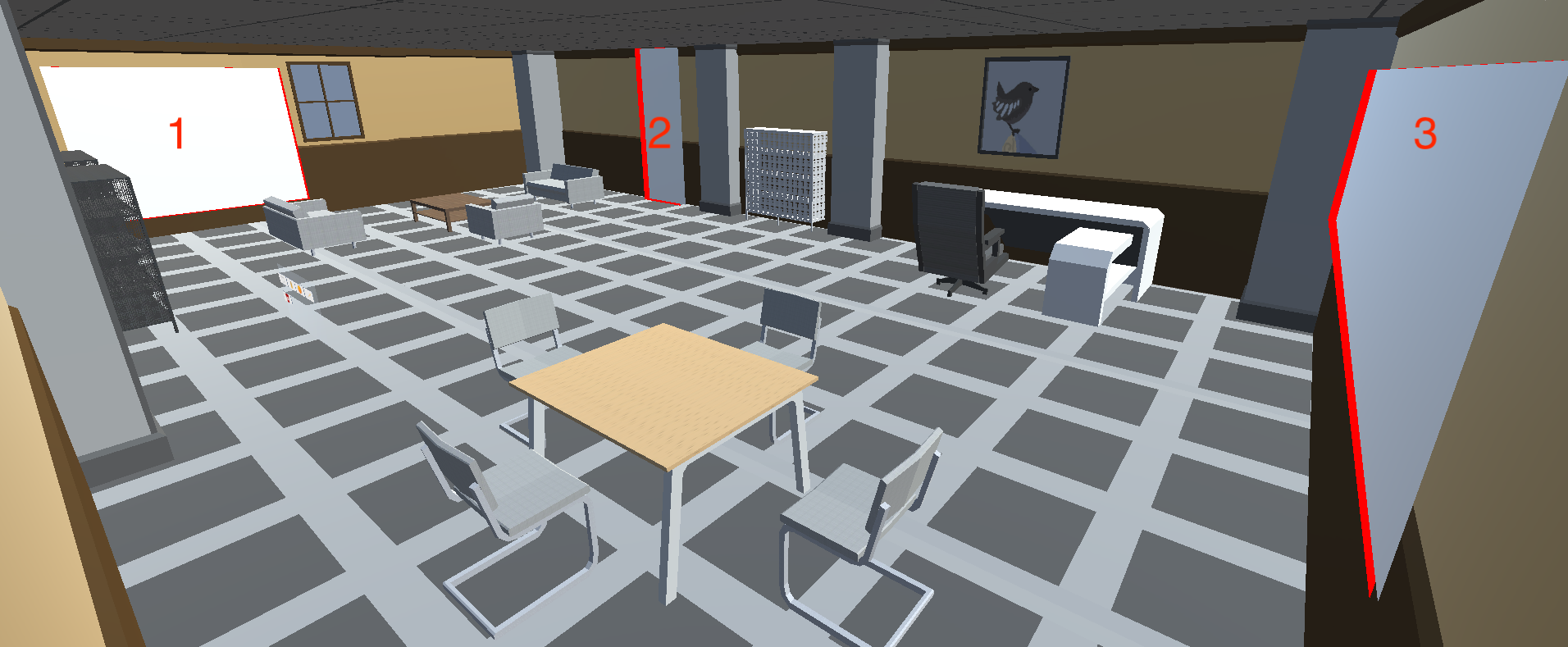}
  \caption{The \textsc{Discrete Choice} technique shown in the Office environment. Users can pick from three shared whiteboard suggestions (Labeled 1, 2, and 3).}
  \label{fig:teaser}
}

\abstract{Augmented Reality (AR) collaboration can benefit from a shared 2D surface, such as a whiteboard. However, many features of each collaborator's physical environment must be considered in order to determine the best placement and shape of the shared surface. We explored the effects of three methods for beginning a collaborative whiteboarding session with varying levels of user control: \textsc{Manual}, \textsc{Discrete choice}, and \textsc{Automatic} by conducting a simulated AR study within Virtual Reality (VR). In the \textsc{Manual} method, users draw their own surfaces directly in the environment until they agree on the placement; in the \textsc{Discrete choice} method, the system provides three options for whiteboard size and location; and in the \textsc{Automatic} method, the system automatically creates a whiteboard that fits within each collaborator’s environment. We evaluate these three conditions in a study in which two collaborators used each method to begin collaboration sessions. After establishing a session, the users worked together to complete an affinity diagramming task using the shared whiteboard. We found that the majority of participants preferred to have direct control during the initialization of a new collaboration session, despite the additional workload induced by the Manual method.
 
} 

\CCScatlist{
  \CCScatTwelve{Human-centered computing}{Human computer interaction (HCI)}{Interaction paradigms}{Mixed / augmented reality};
  \CCScatTwelve{Human-centered computing}{Collaborative and social computing}{Empirical studies in collaborative and social computing}{};
}

\begin{document}


\maketitle

\section{Introduction}
Synchronous collaboration software such as Zoom \footnote{\url{https://www.zoom.com/}}, Microsoft Teams \footnote{\url{https://www.microsoft.com/en-us/microsoft-teams/group-chat-software-b}}, and Webex \footnote{\url{https://www.webex.com}} are widely used by the general population today to conduct business and collaborate with other individuals from anywhere in the world. This existing collaboration software allows users to work together on a shared, virtual whiteboard where any user in the session can contribute, edit, and reference the content on the whiteboard. Shared whiteboards are useful in that they allow users to visualize complex topics being discussed in a meeting that would otherwise be confusing when only being verbally discussed, such as financial data or math equations. People often use whiteboards to sketch, brainstorm ideas, and to group or cluster ideas \cite{cherubini_lets_2007, walny_visual_2011, abascal_using_2015}. 

As immersive technologies such as AR continue to grow in popularity and develop technologically, there will be a desire from the user base to utilize immersive technology to collaborate with other individuals. AR has clear benefits for remote collaboration such as allowing remote collaborators to work alongside one another as if they were present in the same location, which can be critical for activities such as ideation and brainstorming \cite{yoon_effect_2019, aseeri_influence_2021}. However, the shift to immersive technologies for collaboration introduces challenges that will need to be solved to maintain a seamless and effective collaboration experience.  

In traditional collaboration applications designed for 2D displays, the shared whiteboard typically serves as an anchor point, with collaborators’ cursors positioned relative to it. In contrast, immersive collaborative applications must account for the physical positions of collaborators in reference to the shared anchor. This becomes particularly challenging in industrial collaboration meetings, where participants often join from vastly different physical environments. Consequently, designing shared virtual elements and positioning collaborators in a manner that makes sense in the diverse physical contexts of all participants is critical. Such alignment steps, although essential in establishing a coherent, shared positional frame of reference for virtual elements across collaborators~\cite{congdon_merging_2018, keshavarzi_optimization_2020, fink_re-locations_2022}, add additional cognitive load to the user.

One specific benefit of AR collaboration is the ability to work on a shared surface such as a whiteboard; however, determining an appropriate location, size, and shape of the whiteboard that works in multiple physical environments is a difficult problem as each collaborator's physical environment could have differing amounts of available space as well as differing room layouts. It could be possible to scan each user's environment and determine compatible characteristics of the whiteboard by analyzing each environment scan holistically, but this would reduce the amount of control the user has over the shared whiteboard that they would be collaborating on.

In order to inform the design of future remote collaboration experiences in AR, we wanted to explore how the level of control during the initialization process of a collaborative whiteboarding session affected the overall experience of collaboration. We report on the findings of a user study in which participants tested three different conditions (\textsc{Manual}, \textsc{Discrete Choice}, and \textsc{Automatic}) to begin an AR collaborative session with another remote user. To the best of our knowledge, little to no research has been conducted regarding the idea of initialization techniques and assistance for collaboration sessions when using AR to collaborate remotely. Through conducting this user study, we sought to answer the following research question: ``\textbf{What is the impact of system-assisted initialization on the
user experience of remote, dyadic shared surface collaboration
in augmented reality?}'' Our findings showed generally that people preferred having control over the location, size, and shape of the whiteboards that they created.

\section{Related Work}
\subsection{Initialization of CSCW Software}
Our literature review did not reveal any work that specifically dealt with the initialization of CSCW software. This area is an important aspect of CSCW software that is under-researched and should be explored further. This is likely because the initialization portion of CSCW software was straightforward and simple for standard, 2D applications whereas AR adds some complexity, such as aligning multiple physically distinct spaces, that requires further research. We hope that this paper will serve as inspiration for future literature regarding the initialization process when spatially collaborating using AR.  

\subsection{Physical Space Alignment}
When users are collaborating remotely using AR technologies, the users' physical environments must be taken into consideration to support collaboration while maintaining spatial awareness among all collaborators. This is a complicated problem to solve, as users are able to collaborate from any environment with differing sizes, shapes, and obstacles that do not match their collaborators' environments. Aligning physical spaces for collaboration has been a topic of interest within the research community over the past several years \cite{ens_revisiting_2019, marques_remote_2022, marques_critical_2022}. This problem has been approached in a variety of ways with the two main approaches being manual alignment and automated alignment. 

The first method proposed to solve the space alignment problem is by manually aligning differing physical spaces. Prior literature has proposed scanning each collaborator's physical space and then manually annotating similar areas within the scan which can then be used to align collaborators \cite{pejsa_room2room_2016, congdon_merging_2018}.

Another common strategy for aligning physical spaces is to create an automated process that can intelligently analyze scans of each collaborator's physical space and find common area overlaps to align the space to maintain spatial awareness between collaborators. Automating the alignment process is desirable as manually finding areas of overlap can be a time consuming and unintuitive process, especially as the number of collaborators increases. 

We see examples of automated alignment work in works by Lehment et al. and Keshavarzi et al. which take in room scans as input and intelligently find common areas between collaborator's spaces \cite{lehment_creating_2014, keshavarzi_optimization_2020}. An interesting thing to note is that this work allows collaborators to move about their environment while reflecting these movements in a natural way within their collaborator's environment. Kim et al. detailed an algorithm that used Object Cluster Registration to intelligently find overlapping shared space between two different rooms\cite{kim_object_2024}. Kim et al. also detailed a redirected walking method that allowed a VR user to appear in an AR user's environment naturally by tweaking the translation gains of the movement of the VR user, thus maintaining spatial awareness \cite{kim_adjusting_2021}. Kang et al. expanded upon the idea of Kim et al.'s previous work by incorporating a neural network that could alter avatar movement to maintain realism \cite{kang_real-time_2023}. Yoon et al. also used neural networks in a similar manner\cite{yoon_placement_2022}.  

Some existing literature details methods in which collaborators appear in the spaces of other collaborators, but are confined to an area or otherwise hindered from moving about the environment in a natural way. This way, the environments themselves do not need to be aligned, but rather a common anchor is established within each collaborator's environment. Herskovitz et al. provide a toolkit capable of displaying collaborators through a portal, world-in-miniature display, or by anchoring them to a common element of both rooms, such as a chair or table \cite{herskovitz_xspace_2022}. The idea of anchoring has also been explored in other works \cite{gronbaek_partially_2023, fink_re-locations_2022, huang_surfshare_2023}. Yang et al. demonstrate a system that also utilizes anchoring to maintain spatial awareness while providing visual guidance to the collaborators regarding where they should stand to avoid unrealistic placements such as in the middle of a table \cite{yang_visual_2024}.

Collaborating around a shared surface such as a whiteboard does not require strict environment alignment. Thus, we used the shared whiteboard as an anchor for the avatars. However, we still must address the problem of initializing the whiteboard with an appropriate location, size, and shape in each environment, in a way that will facilitate collaboration.

\section{Design Process of Initialization Techniques}
To explore how the level of user control over the initialization of a shared whiteboard affected their collaboration, we designed three initialization techniques ranging from fully manual control to fully automatic initialization. To avoid the limitations of current AR systems and the dynamic nature of real environments, we prototyped and evaluated these techniques in VR. 

All the techniques result in a shared whiteboard among a group of collaborators, with each collaborator seeing the whiteboard in their own virtual environment. Once the whiteboard is established, we show an avatar for each collaborator that reflects their position relative to the shared whiteboard in their respective environment. Doing this allows all collaborators to have a shared understanding of the other collaborators' position, gaze, and pointing gestures toward elements on the shared whiteboard. To keep things simple, we also assumed that only two users were collaborating during our prototyping and evaluation.

\subsection{Manual Technique}
The goal of the manual initialization technique was that each user should be able to place a whiteboard anywhere in their own environment. However, the whiteboards would need to have the same size and shape to be used as shared surfaces. 

Our first prototype was the \textit{Serial Manual} technique. The technique began by having User 1 create a whiteboard in their environment that they felt was in an optimal location and was large enough to complete the given task. This was done by pointing the ray emanating from the controller at a point on the wall, holding the primary trigger on either controller and then dragging the ray to another point on the wall and releasing the trigger. This would create a rectangle with the upper, left corner defined by the initial trigger press and the lower, right corner by the trigger release. User 1 then confirmed the whiteboard and sent it to User 2 by pointing at the whiteboard with the controller ray and pressing the primary button. User 2 would then place the whiteboard in their desired location by pointing the ray at the whiteboard and holding the grip on the side of the controller. Once User 2 placed the whiteboard in the desired area, they would then resize the board so that it would fit in the area by pointing the ray at the whiteboard and moving the joystick up and down to uniformly scale the whiteboard. User 2 would then pass the whiteboard back to User 1 where they could relocate or resize the whiteboard again. This process repeated until both users were happy with the location and size of the whiteboard. 

After some initial pilot testing, we found that the \textit{Serial Manual} technique was cumbersome to use and took a long time to allow users to begin collaborating, which we felt would ultimately lead to user frustration.

\begin{figure}[ht]
    \centering
    \includegraphics[width=8cm]{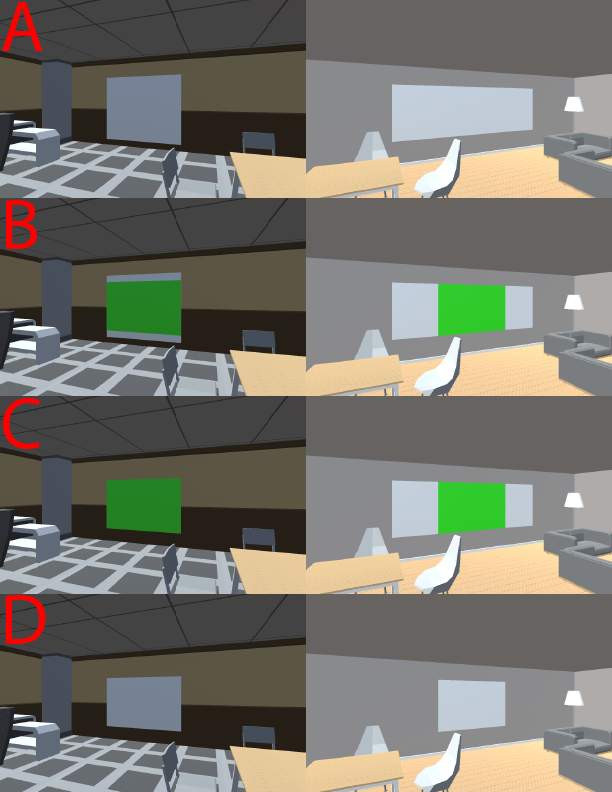}
    \caption{The process that users follow in the \textsc{Manual} condition. A.) User 1 and User 2 each create a whiteboard in their physical environment. B.) Users 1 and 2 are shown the ``Overlap'' (Represented as the green outline) between their two whiteboards. C.) User 1 alters the size of their whiteboard. The overlap is updated accordingly. D.) Both users confirm the whiteboard and begin collaborating with the same shared whiteboard.}
    \label{fig:ManualDiagram}
\end{figure}

Thus, our second prototype was the \textit{Parallel Manual} technique. In this technique, both collaborators created whiteboards in parallel in their respective environments, using the same control scheme described above. After both users created a whiteboard, the system determined their overlap based on the minimum width and height of the two individual whiteboards. Both users were then shown an outline of this size, centered at the location of their original whiteboard, representing the current proposal for the shared whiteboard. Users were free to redraw their individual whiteboards as many times as necessary to get the desired overlapping, shared whiteboard size. Once both users were happy with the size of the shared whiteboard, each user confirmed the shared whiteboard by pointing the controller ray at the whiteboard outline and pressing the primary button. We demonstrate this process in Figure \ref{fig:ManualDiagram}. It is worth noting that the location of either collaborator's whiteboard did not affect the other collaborator's whiteboard position. Both collaborators could place the whiteboard wherever they wanted within their respective environments. 

In our preliminary testing, we found the \textit{Parallel Manual} technique to be much quicker and overall more efficient in establishing a common shared whiteboard that was compatible with both user's individual environments. Thus, we included it in the user study as the \textsc{Manual} condition.

\subsection{C-SAW}
Before developing system-assisted initialization techniques, we developed some rules for the system to follow when determining whiteboard location, size, and shape. We propose C-SAW (Collaborative Surface Algorithm for Whiteboarding), which determines candidate whiteboard placements and sizes based on the constraints of the physical environments of a group of users. 
Because we wanted to explore the effectiveness of this approach, C-SAW was not implemented to be intelligent and autonomous. Instead, we developed a set of heuristics humans could use to create suggested whiteboards depending on the specific combination of environments that each collaborator was in. They are as follows:

\begin{figure}[ht]
    \centering
    \includegraphics[width=8cm]{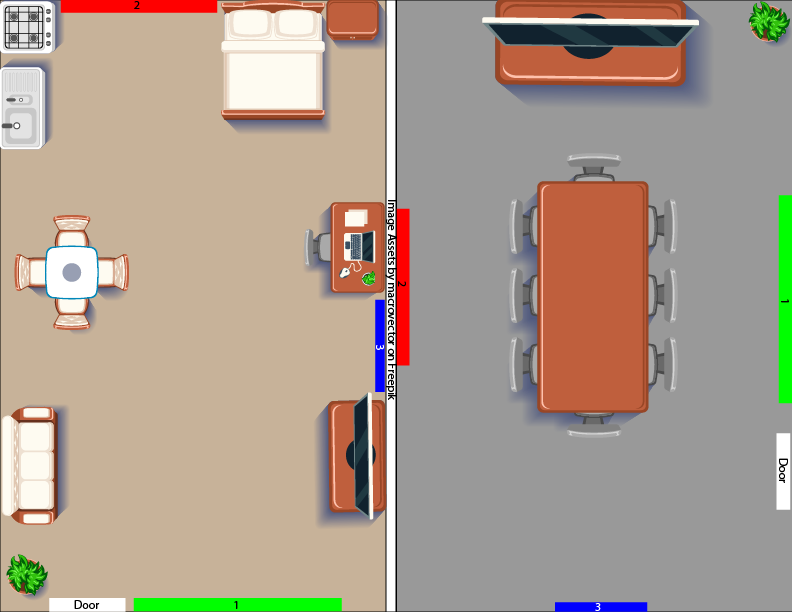}
    \caption{An example of C-SAW (Specifically, C-SAW working with the \textsc{Discrete Choice} technique) being applied to two collaborator's respective environments. The collaborators are offered three whiteboard suggestions that fit the constraints of both environments (Shown as the red, green, and blue numbered rectangles).}
    \label{fig:CSAWDiscreteChoiceDiagram}
\end{figure}

\begin{enumerate}
    \item The user should be able to reach the whiteboard
    \item The whiteboard should not be blocking something important (e.g., window, door, etc.)
    \item The whiteboard should be placed in relatively open space on a wall
    \item The whiteboard should not be so narrow or short that content cards do not fit completely on the board
    \item The whiteboard should have enough free space in front of it for the user to be able to view the entire whiteboard at once
\end{enumerate}

We show an example of C-SAW being applied to two collaborators' environments in Figure \ref{fig:CSAWDiscreteChoiceDiagram}.

To avoid bias in the size and location of the whiteboard in the study environments, a researcher who was not affiliated with the project was consulted to provide the sizes and locations of the whiteboards in the environments used throughout the user study. During the selection process, the researcher followed the rules of C-SAW to select whiteboard sizes and locations.  

\subsection{Automatic}
The \textsc{Automatic} initialization technique is the fully automated technique that uses C-SAW to look at each collaborator's respective environment holistically. There is no user interaction in this condition. Each collaborator joins the session, and a single shared whiteboard that is compatible with both collaborators' environments is given to them so that collaboration can begin immediately. The whiteboard provided by the \textsc{Automatic} technique is the one that best follows the rules of C-SAW (based on the analysis of the human researcher), meaning the given whiteboard was often in an area with few objects surrounding the whiteboard, as well as being on a wall that had lots of open space. 

\subsection{Discrete Choice}
In addition to \textsc{Manual} and \textsc{Automatic}, we wanted to provide a middle-ground technique. The \textsc{Discrete Choice} technique uses C-SAW to holistically analyze both collaborators' environments and provide three suggestions for whiteboards whose sizes, shapes, and locations are respective of the constraints of each collaborator's environment. After viewing the suggestions, the two collaborators work together to come to an agreement on which whiteboard suggestion to choose. Each user would point their controller's ray at a suggestion and pull the trigger on the controller to make their selection. 

If both collaborators selected the same suggestion, the whiteboard turned green for both users to confirm much like they did in the \textsc{Manual} condition. However, if they selected different suggestions, then the user's selected suggestion was shown in a yellow color while their partner's suggestion was shown in a blue color. Collaborators were free to change their selection before confirming the suggestion as many times as necessary. After each user confirmed the mutual suggestion, the users could begin collaborating.

\subsection{Synchronizing Spatial Awareness}
Once participants created or selected a shared whiteboard for each respective condition each user was then able to see a virtual representation of the other user in their own environment relative to the whiteboard. Each user's created whiteboard acts as the ``anchor'' for the avatar. As an example of this, if User 1 is two meters from the whiteboard in their environment, then User 1's avatar in User 2's environment is also two meters away from the other user's whiteboard. Additionally, if User 1 points at a sticky note in the bottom, left corner of their whiteboard, their avatar in User 2's environment is also pointing at the same sticky note located on User 2's whiteboard. 

\begin{figure}[ht]
    \centering
    \includegraphics[width=8cm]{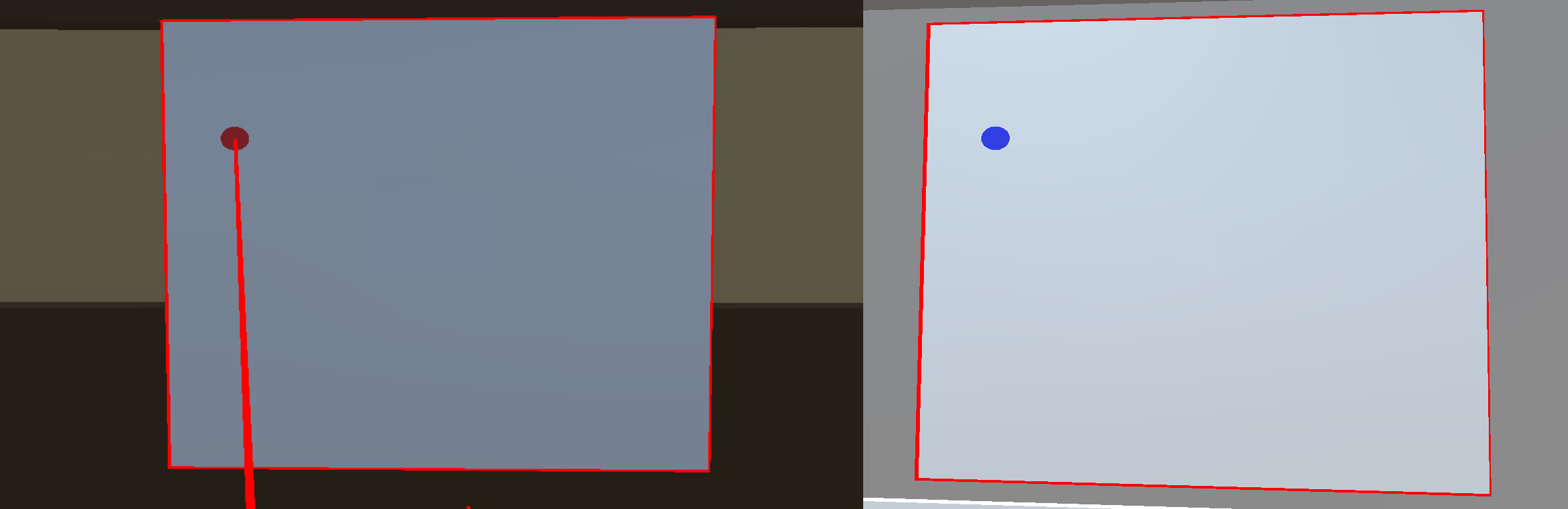}
    \caption{User 1 pointing their laser pointer at the whiteboard in their environment (Left). User 1's laser pointer's position reflected on User 2's whiteboard (Right).}
    \label{fig:laser}
\end{figure}

It is also at this point where each user has a ``laser pointer'' that they can use to point at sticky notes on their respective whiteboard from a distance. Similar to how the avatars work, the intersection of each user's laser pointer with their whiteboard is reflected on the opposite user's whiteboard. User 1 pointing at a note on their whiteboard will be reflected on User 2's whiteboard and vice versa. We show an example of this in Figure \ref{fig:laser}. 

\section{Experiment}
\subsection{Goals}
The goal of this study was to understand how the experience of collaboration was affected by varying the level of control, or, conversely, the amount of assistance, provided to the user during the initial setup process of a shared whiteboard.

Specifically, we examined the following hypotheses: 

\begin{itemize}
    \item H1: Users will prefer the \textsc{Discrete Choice} condition because of its balance between ease of initialization and flexibility.
    \item H2: Users will least prefer the \textsc{Automatic} condition due to its lack of choice in location.
    \item H3: Users will be more frustrated while utilizing the \textsc{Automatic} condition
\end{itemize}

\subsection{Experimental Design}
\begin{figure}[ht]
    \centering
    \includegraphics[width=8cm]{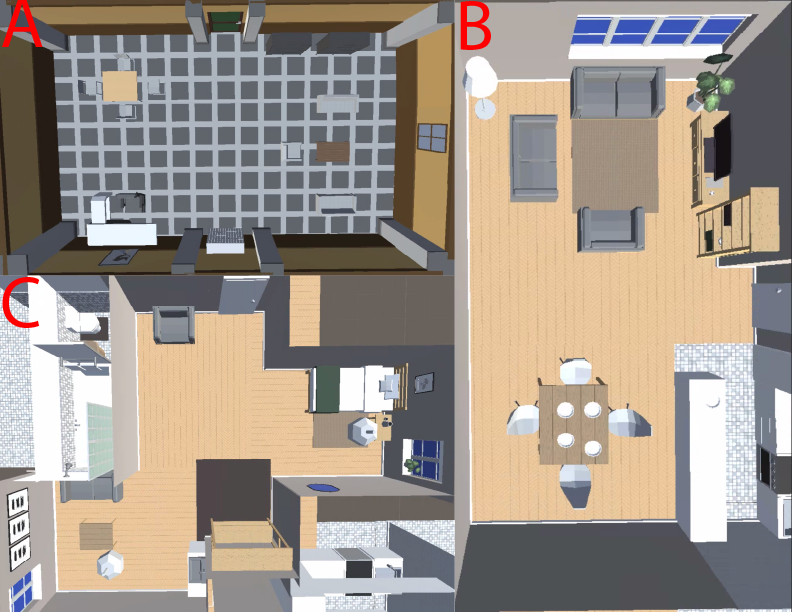}
    \caption{A top-down view of each of the three study environments. A.) Office - B.) Kitchen - C.) Bedroom}
    \label{fig:envtopdown}
\end{figure}

The within-subjects study had one primary independent variable: initialization technique (\textsc{Manual}, \textsc{Discrete Choice}, and \textsc{Automatic}). We also varied each user's environment (Office, Kitchen, Bedroom (shown in Figure \ref{fig:envtopdown})) and the dataset used in the collaboration task (Sports, Food, History) in order to test each technique in a variety of scenarios. The combination of technique and environment were counterbalanced using a balanced Latin square to ensure that there were no environmental biases. Pairs of participants were always in different virtual environments for a particular trial, and each user saw each environment only once. Each participant pair was located in the same physical space throughout the study. Each participant was placed on opposite sides of the room with a barrier placed between them. Participants were free to speak aloud to one another as if they were using online voice communication. 

Multiple dependent variables were collected. We measured user preference for the techniques; time to complete initialization; the size, shape, and location of participant-created whiteboards; and the participant responses to the post-study interview questions. We administered the NASA-TLX\cite{hart_development_1988} and the UEQ\cite{holzinger_construction_2008} in order to gauge how the workload and user experience of collaboration were affected by the initialization technique. We also administered the SUS questionnaire\cite{brooke_sus_nodate} after both the \textsc{Manual} and \textsc{Discrete Choice} conditions to gauge the overall usability of the initialization technique. We did not administer the SUS questionnaire after the participant pair completed the \textsc{Automatic} condition considering that participants did not have to do anything in order to establish the shared whiteboard. 

\subsection{Apparatus}
Each participant used a Meta Quest 2 Head-Worn Display (HWD) and two Meta Touch controllers. The headset has an 1832 x 1920 resolution per eye with a refresh rate of up to 90 Hz. \footnote{https://www.meta.com/quest/products/quest-2/tech-specs/} We also used two PCs running Windows 10 to run the VR applications. One computer had an i9-11900F CPU with an RTX 3080 GPU and 32GB of RAM. The other PC had an i9-12900K CPU along with an RTX 3070 ti GPU and 32GB of memory. We also used two Meta Link cables to connect the HWDs to the PCs. The VR study software was built with Unity version 2020.3.35f1 and was networked with Photon PUN 2 to synchronize avatar positions and user interactions.

\subsection{Task}
In each trial, participants completed two main tasks. The first task involved establishing a shared whiteboard with the other participant. The method through which this was accomplished varied based on the condition that the participant pair was testing.

In the second task, participants worked together to complete an affinity diagramming task. Affinity diagramming is defined by Lucero as, ``a technique used to externalize, make sense of, and organize large amounts of unstructured, far-ranging, and seemingly dissimilar qualitative data\cite{abascal_using_2015}.'' The second task involved organizing virtual sticky notes on the shared whiteboard until the participant pair was satisfied.  

We used six datasets (sets of sticky notes) throughout the study. The first three datasets were used as training data for each of the three interaction techniques and were not included in any of the data analyses in subsequent sections of this paper. The other three sets of data were used in the actual study conditions. These datasets dealt with food, history, and sports. We chose the food and sports datasets as they were general enough that participants could work together to create multiple different groups with the given data. The history dataset allowed us to give a specific grouping directive to the participant pairs (ordering the historical events chronologically). The food and sports sticky notes had images of various food items and sports equipment, while the history set was text-based and featured various events throughout the world's history. The food and history datasets had 20 entries while the sports dataset only had 10 entries. 

\begin{figure}[ht]
    \centering
    \includegraphics[width=8cm]{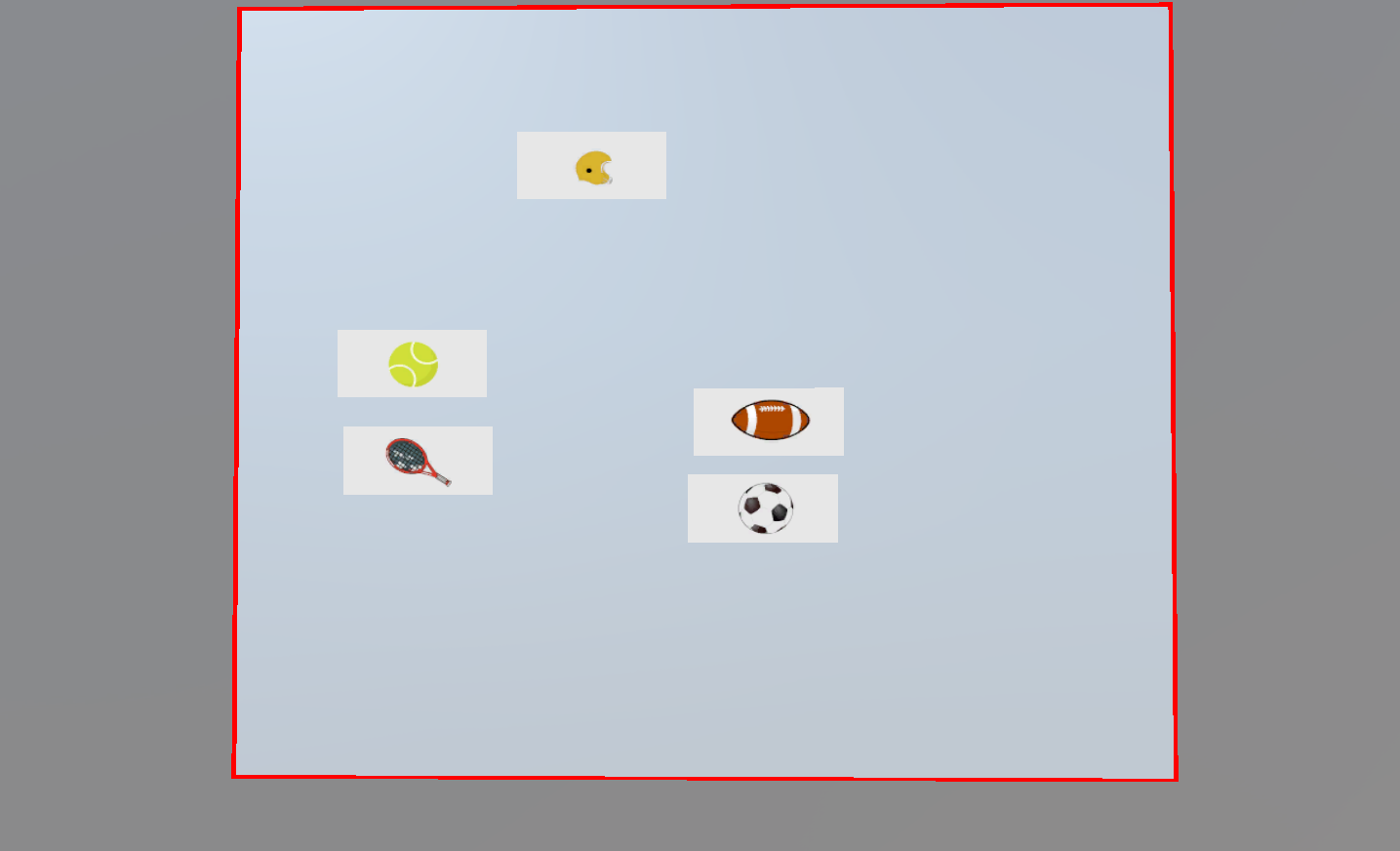}
    \caption{An affinity diagramming task in progress.}
    \label{fig:affinitydiagramming}
\end{figure}

Each participant in a pair received half of the items in a dataset at the beginning of the affinity diagramming task, to ensure that both users would participate equally---participants could only move items belonging to them. The pair were free to discuss organization strategies between themselves. An affinity diagramming task in progress with the sports dataset is shown in Figure \ref{fig:affinitydiagramming}.

When spawning into the study environment, each participant was given their half of the notes on a ``note holder'', a rectangular surface that followed them as they moved around the study environment. The holder was positioned below their waist so that their vision would not be occluded by the holder. The holder was pitched at a 27\degree angle relative to the user to allow them to easily view all the note cards.

Participants could interact with the dataset items by pointing their controller's ray at one of the items and pulling and holding the trigger. Once the participant was holding an item, they could then point their ray at the shared whiteboard where they would be shown an outline of the item's position on the whiteboard if they released the trigger at that moment. Releasing the trigger placed the item on the whiteboard. Participants could then move the item again using the same approach. Participants could also point their controller's ray at the floor and pull the grip on the controller to teleport around the environment.

\subsection{Procedure}
Upon arriving at the laboratory, participant pairs were welcomed and introduced to one another before beginning the study. Before participating in the study, participant pairs completed a pre-study questionnaire where demographics about each of the participants were gathered.

After taking care of the pre-study documents, participants were formally introduced to the study. We began by showing participants a brief pre-study presentation that introduced the concept of AR whiteboards and some of the challenges that whiteboards of this kind could introduce.

After the presentation, participants were introduced to the Meta Quest 2 and the accompanying Meta Touch Controllers. Participants were shown the different buttons and triggers that they would need to use. Participants were then shown how to wear and adjust the fit of the HWD to a comfortable setting. Participants were then instructed to put the HWD on and adjust it so that the fit was comfortable and the content in the headset was legible and easy to see. 

Next, participants began training for the first interaction technique they would be utilizing. After training, the participants would then complete the actual trial condition using the interaction technique they had just trained for. This process was repeated for each of the three interaction techniques. 

After completing the trial for a given condition, participants completed surveys regarding both the experience of establishing a shared whiteboard as well as how the initialization technique affected the overall experience of grouping the sticky notes. Participants only completed one trial for each of the techniques. 

After completing all three interaction techniques, participants participated in a joint post-study interview in which they were asked questions regarding their experience establishing their shared whiteboards for each of the three conditions; their verbal responses were recorded. Participants were asked questions about their most and least preferred conditions, the condition that produced the most ideal whiteboard, negotiation strategies during initialization between them and their partner, the pros and cons of each condition, and their thoughts on the importance of whiteboard size, shape, and location. After concluding the post-study interview, we thanked both participants for their time and concluded the study session.

\subsection{Participants}
After receiving IRB approval from our university, we recruited 36 participants (27 Male, 9 Female) for a total of 18 pairs of participants. Participants were 18 years of age or older, with normal or corrected vision (i.e., contacts). Participants who wore glasses were not included in the study. Participants were recruited from various computer science and Human-Computer Interaction email lists. Participants were between the ages of 19 and 35 (M=23.08, SD=4.17)

Concerning prior experience, 16 of our participants had never used AR before, 12 had used AR 1-3 times, 4 had used AR 5-10 times, and 4 had used AR more than 10 times. 5 had never used VR, 18 participants had used VR 1-3 times, 5 participants had used VR 5-10 times, and 8 people had used VR more than 10 times. 12 participants said that they had never met their study partner, 3 participants said that they were acquaintances with their study partner, and 21 people said that they were friends with their study partner. 

\section{Results}
\subsection{Technique Preference}
During the post-study interview, participants were polled on their favorite and least favorite initialization techniques.

\begin{figure}[ht]
    \centering
    \includegraphics[width=8cm]{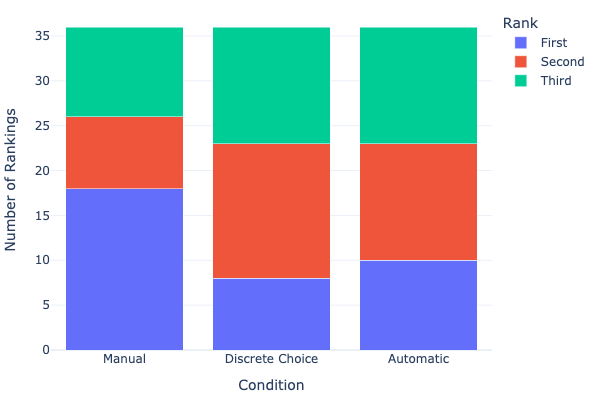}
    \caption{Participant rankings of each of the initialization techniques}
    \label{fig:FrequencyPlot}
    \vspace{-0.25cm}
\end{figure}

As Figure \ref{fig:FrequencyPlot} shows, half of the participants ranked the manual condition as their first choice. Eight ranked the \textsc{Discrete Choice} technique first, while ten ranked the \textsc{Automatic} condition as their favorite. We found little difference between the rankings of each of the initialization techniques. A chi-square test (F(4) = 7.33,p= 0.12) revealed no significant differences in the ranking of the techniques.

\subsection{Questionnaire Results}

We measured average SUS scores of 80.42 and 84.31 for the \textsc{Manual} and \textsc{Discrete Choice} conditions, respectively.

We used unweighted NASA-TLX scores in our analysis. The overall workload scores were 3.61, 3.57, and 3.76 for the \textsc{Manual}, \textsc{Discrete Choice}, and \textsc{Automatic} conditions, respectively. The NASA-TLX subscale values had very similar results across all three of the study conditions. The Mental Demand, Physical Demand, Temporal Demand, Effort, and Frustration subscales all had low values that never exceeded \textbf{3.6}. Likewise, the performance subscale never dipped below \textbf{8.3} for any of the conditions. 

The results from the UEQ responses were scored by summing each of the results for a participant in each of the six scales (i.e., Attractiveness, Perspicuity, Efficiency, Dependability, Stimulation, and Novelty) for each participant for each condition. We found that each of the scales produced very similar scores across all three conditions with no scale scoring lower than 3.31 and no higher than 5.58.

\subsection{Time to Complete Initialization}
After participants completed the initialization process, we marked the exact time that the process completed. We found that on average it took participants 45 seconds to complete the initialization for the \textsc{Manual} condition. In addition, we found that it took participants an average of 26 seconds to complete initialization in the \textsc{Discrete Choice} condition. It should be noted that average times were not calculated for the \textsc{Automatic} condition as there was no input required from the participant for initialization.

\subsection{Data and task Influence on Whiteboard Characteristics}
\begin{figure}[ht]
    \centering
    \includegraphics[width=8cm]{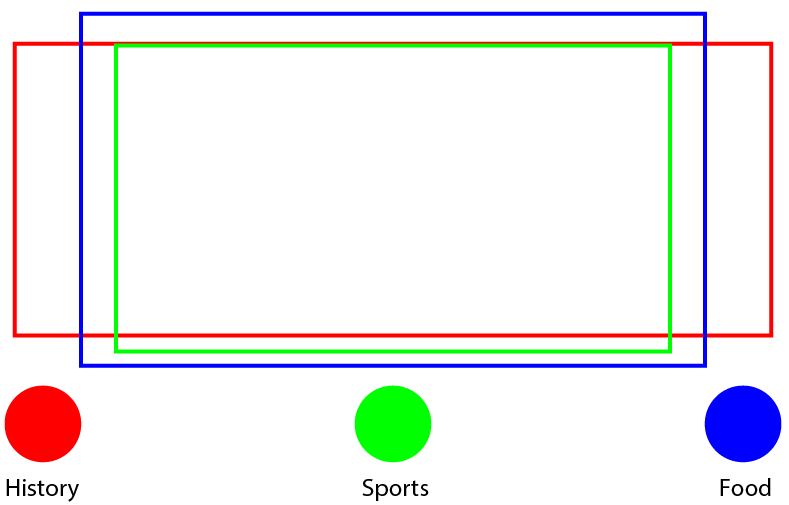}
    \caption{Average shapes and relative sizes of whiteboards created by the participants in the \textsc{Manual} condition.}
    \label{fig:averagewhiteboardsbytask}
    \vspace{-0.25cm}
\end{figure}

Participants exhibited behavior that suggested that the dataset directly influenced the characteristics of the whiteboard participants either created or selected. Quantitatively, we observed that the average scale of the whiteboards created by participants in the manual condition increased as the number of items to group increased. Specifically, the average scale of whiteboards created with the sports dataset (with ten items) were smaller (3.5m x 1.93m) compared to those created when using the 20-item food dataset (3.94m x 2.22m). We show the average sizes of the whiteboards overlaid on one another in Figure \ref{fig:averagewhiteboardsbytask}.

However, the amount of data alone does not influence the characteristics of created whiteboards. Participant pairs were instructed to work together to order the history dataset chronologically. As a result, we found that the average whiteboard for this dataset was wider and less tall (4.78m x 1.84m) compared to the whiteboards that used the food and sports datasets. Participants created thin and long whiteboards that resembled a timeline as a result of being given a timeline task. We also observed that 10 out of 18 participant pairs chose a whiteboard with a larger aspect ratio during the \textsc{Discrete Choice} condition. 

\section{Discussion}
The study was designed to answer the following research question: \textbf{``What is the impact of system-assisted initialization on the user experience of remote, dyadic shared surface collaboration in augmented reality?''}

\subsection{Tradeoffs Among Initialization Techniques} \label{ParticipantPreference}
Prior to conducting the study, we hypothesized that participants would prefer the \textsc{Discrete Choice} condition because of the balance it provided between ease of initialization and flexibility \textbf{(H1)}. We also hypothesized that participants would least prefer the \textsc{Automatic} condition because of the lack of choice in whiteboard location \textbf{(H2)}. We found no significant difference between the technique rankings. \textsc{Discrete Choice} actually had the fewest first place rankings, while \textsc{Manual} had the highest. Thus, the results do not support \textbf{H1} or \textbf{H2}. We also found that participants were not frustrated with the \textsc{Automatic} condition which does not support \textbf{H3}.

The interview responses shed some light on why participants ranked the conditions the way that they did. Half of the participants stated that they preferred the \textsc{Manual} technique because of the control and freedom that the condition provided them. P9 said, \textbf{``I like the one where you could draw your own the best because, well, for the task...It was better to be able to customize the size of your own board. Like we wanted a longer timeline ..."}

Participants also mentioned liking the freedom to create whiteboards specifically tailored for the given dataset in the \textsc{Manual} condition. Whiteboards became wider and less tall with the history dataset (Timeline task), and larger with the datasets containing more items, as described in Section 5.3. Some participants told us during the post-study interviews that they attempted to create a whiteboard that resembled a timeline.  P29 commented on the whiteboard shape being influenced by the task at hand, \textbf{``... let's say you were to put something like as a long rectangle or something like that. A lot of people like consider that to be like chronological or like a timeline or something.''}

On the other hand, a large number of the participants found the workflow in \textsc{Manual} to be unintuitive and difficult to use. Nearly all of the participants struggled initially with the technique, even if they did not explicitly say so during the post-study interview. We found that on average a participant pair together created 10 (M=10.44, SD=7.76) different whiteboards prior to agreeing on a whiteboard which suggests difficulty in using the technique. During the post-study interview, seven participants explicitly mentioned the complexity of the \textsc{Manual} condition as a reason for ranking it as their least favorite. P3 said, \textbf{"Probably the last one [\textsc{Manual}]. The overlapping of the whiteboard thing feels a little complicated. I mean, I didn't have any trouble with it, but I feel like if more people were doing that, I feel like it takes a little bit more time. It's, you know, not a super huge amount of advantage..."}

Similarly, five participants mentioned that they did not like the \textsc{Manual} condition because of the time required to establish a session with the other collaborator. P1 said, \textbf{``…I guess least favorite would be the first one [\textsc{Manual}]. Just because of how many steps you have to do to get a shared collaborative space [Whiteboard]."} On average, it took participant pairs 45 seconds to begin collaborating in the \textsc{Manual} condition.

While we may not have found a statistically significant difference in the participant rankings of the three conditions, the participant ranking data, as well as comments made to us during the post study interview allow us to conclude that collaborators do prefer to have explicit control over the location, size, and shape of shared whiteboards, but that our design needs further improvements in efficiency and understandability.

It is also worth noting that the majority of the technical issues during the study occurred during the \textsc{Manual} condition. There were two main technical issues encountered during the condition. The first issue was that occasionally the created whiteboards would not be passed to the other respective collaborator's Unity instance. This was likely caused by network instability and a restart of the study software would resolve the issue. The other issue occurred when participants were attempting to create their whiteboard on a wall. Participants would sometimes create whiteboards so large that the corner of the whiteboard would begin to encroach into the corner of the adjacent wall which would prevent the whiteboard from being created. This led to visible frustration from the participants. These two issues may have influenced participant perception of the \textsc{Manual} condition.

Regarding the \textsc{Discrete Choice} condition, a small subset of participants seemed to like having multiple options to choose from within their respective environments. In the post-study interview, five participants explicitly mentioned liking the fact that the \textsc{Discrete Choice} condition provided them with options for whiteboard size and location. P11 said, \textbf{``…the \textsc{[Discrete Choice]} one, I think that is the best. It still gives you to take some decision where you want to ... You can actually just pick the right position, and you don't have to do a lot ... to draw it on the walls.''} 

\begin{figure}[ht]
    \centering
    \includegraphics[width=8cm]{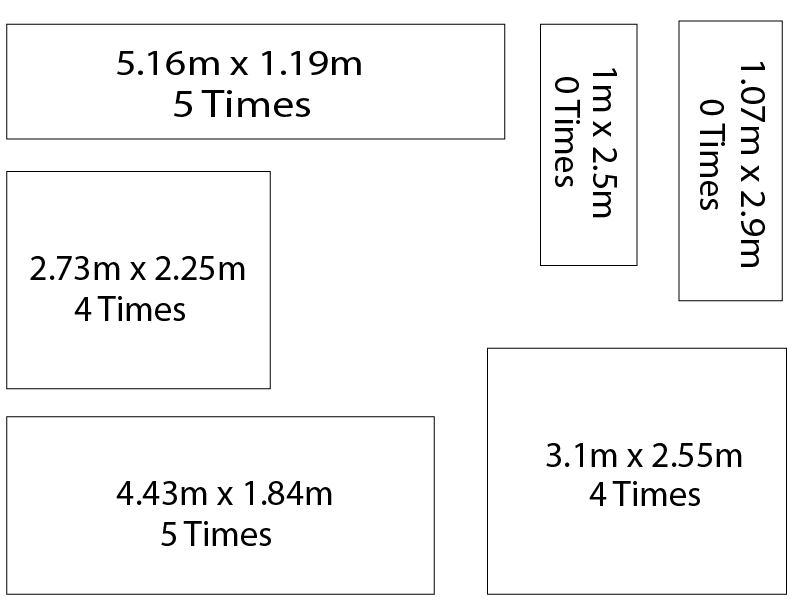}
    \caption{The whiteboard suggestions selected by each participant pair during the \textsc{Discrete Choice} condition.}
    \label{fig:selectionsinsuggested}
\end{figure}

One aspect of the \textsc{Discrete Choice} condition that participants sometimes struggled with was the perceived uselessness of some of the whiteboard suggestions provided during the condition. We attempted to provide a variety of whiteboards that had different shapes and sizes while still conforming to the requirements of the individual environments that each participant was in. However, we found that no participants ever selected the whiteboard choices whose height was greater than their width (see Figure \ref{fig:selectionsinsuggested}). We expected that some participant pairs would prefer to use such a whiteboard during the timeline task to order events from the top to the bottom. During the post-study interview, some participants said they could never see a potential scenario of using the vertical whiteboard. P28 said, \textbf{``... if the vertical board was horizontal, like the one that was next to it [horizontal board] was still better than having the vertical one that went like straight to the ground.''} 

During the interview, 10 of the 36 participants mentioned that they believed that the \textsc{Discrete Choice} condition took the longest to begin collaborating. Some did not like the concept of choosing suggestions at all, while others had issues agreeing on a common, shared whiteboard suggestion to use. One participant pair in particular (P33/34), actually argued over which suggestion to use and it took them 77 seconds to begin collaboration (compared to the average time in the \textsc{Discrete Choice} condition of 26 seconds). It could be that participants would find it less frustrating to exert more effort in the \textsc{Manual} condition or rely on the \textsc{Automatic} condition to provide the ideal whiteboard rather than debate with their partner over which suggestion to use. 

With the food and sports datasets, participants told us that during the \textsc{Manual} and \textsc{Discrete Choice} conditions, they often just went with the largest whiteboard they could either create or select. When asked which condition produced the most ideal whiteboard, P4 said, \textbf{``Probably manual, but that's just because we can make it as big as possible."} P4's collaborator P3 said this when asked if the suggestions in the \textsc{Discrete Choice} condition were similar to what they would have created themselves: \textbf{``I think they differ just because I probably would have gone for the biggest surface area possible and just used whichever part of it I needed."} This suggests that it might be attractive to allow users to resize the suggestions provided by the \textsc{Discrete Choice} technique as users could find the ``best" suggestion and then tweak the dimensions of the whiteboard to fit their exact environment. It also suggests that whiteboard size should perhaps be given the highest weight in the C-SAW algorithm.  

Because the \textsc{Automatic} condition had the least amount of user involvement, participants ultimately had the least to say regarding this condition. One aspect that participants seemed to like was having the shared whiteboard ready for collaboration with no input or work required by them. Ten participants made comments regarding the convenience and simplicity of the \textsc{Automatic} condition. P7 said, \textbf{``I think the auto is definitely the most efficient, like for just getting right into it, right into actually doing the task."}

Participants were satisfied overall with the whiteboard provided by the \textsc{Automatic} condition, with all 36 participants mentioning during the post-study interview that they were happy with the whiteboard provided. Some participants did not care for the location of the whiteboard with 3 of 36 participants explicitly mentioning poor location during the post-study interview. Otherwise, any other negative feedback regarding the \textsc{Automatic} condition was related to the characteristics of the condition itself (i.e. having no control over the whiteboard for collaboration). 

On the other hand, participants also said that the \textsc{Automatic} condition was not flexible at all and that the whiteboard chosen by the technique sometimes had issues, such as being partially occluded by furniture or other objects within the environment. Eleven participants explicitly mentioned one or both of these issues during the post-study interview. P30 said \textsc{Automatic} was their least favorite condition, ``\textbf{because like they only gave me one option. But with the second one[\textsc{Discrete Choice}], they gave me three options.}'' P31 commented, \textbf{``I felt like it [the whiteboard] was placed in an area that was near furniture and ... it just felt like the furniture was in the way of viewing the whiteboard sometimes so you have to get extra close to it.''} Thus, future C-SAW algorithms need to be intelligent in order to make appropriate selections of whiteboard location, size, and shape. It may be worth considering some kind of ``manual override'' option in the \textsc{Discrete Choice} and \textsc{Automatic} conditions where the user can override the suggestion and make their own whiteboard. In general, our data suggest that participants liked automatically being able to collaborate with no input or effort required from them, as long as the algorithms selecting the whiteboard location, size, and shape work well.

\subsection{Design Considerations for Collaborative Session Initialization}
When designing the initialization experience for a collaboration session, there are some design considerations that are apparent from the results of our study. 

First, we suggest that the best initialization technique for shared surfaces could be a combination of the three conditions tested in the study. Our data showed that both manual control and intelligent system suggestions could be effective and preferred. We envision a new technique that would use an intelligent C-SAW algorithm to offer one whiteboard suggestion based on the task or data the collaborators were working on. If collaborators do not like the initial whiteboard, they could then select from other intelligent suggestions or modify any of the suggestions manually. This would give collaborators the ultimate flexibility to get a whiteboard to provide the best experience for collaboration. We propose the prototyping and evaluation of this approach as future work.

The second design consideration for collaborative initialization techniques is keeping all collaborators aware of the actions and data elements of other collaborators in the session. We observed that 34 of our 36 participants placed all of their data onto the shared whiteboard prior to grouping them because it was easier to discuss grouping strategies when all the data was already visible to both users. Future AR collaboration systems should implement systems that allow users to see the data to be organized or worked within a session to keep all collaborators in the loop without making extra work for them. 

Awareness of collaborator actions is also an important issue. In the \textsc{Manual} condition, participants were unable to see the size and shape of the whiteboard that their collaborator created. Instead, they could only see the overlap after both users had created their whiteboards. Three participants explicitly mentioned how useful it would be to be able to see the dimensions and shape of the whiteboard that they were creating. P2 commented, \textbf{``When you're trying to match up a whiteboard ... it would be better if you ... see what the other person's drawing [Whiteboard] is like."} Keeping both collaborators in the loop would aid in the negotiation of the shared surface, reducing the number of times the whiteboards would need to be edited and saving time overall. 

\section{Limitations \& Future Work}
This work had several limitations. First, our study was conducted by simulating AR within VR. A study similar to ours should be conducted with an AR HWD to see if the results from our study change at all.

Our study and whiteboard collaboration software was designed to simulate two individuals working together at a physical whiteboard, thus we added the requirement that whiteboards needed to be exactly the same and fit within each user's respective environment. Therefore, our system did not include features that you might find in digital whiteboard collaboration software, such as ``infinite'' scrollable whiteboards. A future study should be conducted that includes the affordances provided by digital whiteboards to see how the results of the study are altered. 

In addition, further refinements to the control scheme in the \textsc{Manual} and \textsc{Discrete Choice} conditions would likely improve participant understanding of those conditions and could influence participant condition preference. 

Finally, our work only evaluated initialization techniques in the context of an affinity diagramming task. Other shared surface tasks, such as presentation, annotation, or sketching, should be studied in the future. 

\section{Conclusion}
Making AR collaboration more user-friendly will lead to increased adoption of AR in industrial settings. This was a major reason we focused on the initialization aspect of a new collaborative session, as a poor setup experience could lead to user frustration and a hesitance to use AR for collaboration on projects within industry. 

In this work, we looked at how different levels of control during the initialization phase of a collaborative whiteboarding session affected the overall collaboration experience. We tested three conditions --- \textsc{Manual}, \textsc{Discrete Choice}, and \textsc{Automatic} --- and found that participants overall preferred having control over the size, shape, and location of whiteboards within their environment. Our qualitative evidence also suggested that users would be satisfied with an \textsc{Automatic} technique as long as the algorithm selecting the whiteboard position, size, and shape was intelligent enough to make a choice that matches the expectations of the users.


\bibliographystyle{abbrv-doi}

\bibliography{whiteboard}
\end{document}